\documentclass[twocolumn,showpacs,superscriptaddress]{revtex4}%
\usepackage{amsfonts}
\usepackage{amsmath}
\usepackage{mathrsfs}
\usepackage{amssymb}
\usepackage{wasysym}
\usepackage{subfigure}
\usepackage{color}
\usepackage{graphicx}
\usepackage{url}
\providecommand{\U}[1]{\protect\rule{.1in}{.1in}}
\begin{document}
\title{Analytic calculation of high order corrections to quantum phase transitions of ultracold Bose gases in bipartite superlattices}
\author{Zhi Lin}
\email{zhilin13@fudan.edu.cn}
\affiliation{Department of Physics and State Key Laboratory of Surface Physics, Fudan University, Shanghai 200433, P.R. China}
\affiliation{Shenzhen Institute of Research and Innovation, The University of Hong Kong, Shenzhen 518063, China}
\author{Wanli Liu}
\email{wanliliu12@fudan.edu.cn}
\affiliation{Department of Physics and State Key Laboratory of Surface Physics, Fudan University, Shanghai 200433, P.R. China}

\begin{abstract}
We clarify some technical issues in the present generalized effective-potential Landau theory (GEPLT) that makes the GEPLT more consistent and complete. Utilizing this clarified GEPLT, we analytically study the quantum phase transitions of ultracold Bose gases in bipartite superlattices at zero termparture. The corresponding quantum phase boundaries are analytically calculated up to the third-order hopping, which are in excellent agreement with the quantum Monte Carlo (QMC) simulations.
\end{abstract}
\pacs{03.75.Hh, 64.70.Tg, 67.85.Hj, 03.75.Lm}
\maketitle

\section{Introduction}
The physics of ultracold Bose gases in optical lattices  has been a hot topic during the last decade \cite{Lewenstein,RMP}. Since the intensities and forms of interaction are highly controllable \cite{RMP,synthetic-gauge-field1,SOC1,synthetic-gauge-field2,SOC2,Eckardt} for ultracold quantum gases, there is great potential to simulate the physics of quantum many-body systems \cite{QS1,QS2,QS3} that include a broad range of fields such as atomic physics, quantum chemistry, high energy physics, cosmology, and condensed matter physics.

As a famous example of simulation in condensed matter physics, the Bose-Hubbard model \cite{BHM1,BHM2} has been realized by loading ultracold bosonic atoms in homogeneous simple cubic lattices \cite{Greiner}. In this system, ultracold bosonic atoms can experience a quantum phase transition from a  Mott insulator (MI) to a superfluid (SF) when the ratio of the hopping amplitude $t$ to the on-site interaction $U$ goes above some critical value. The MI-SF transition can be detected directly in experiments via time-of-flight technique \cite{Greiner}. Recently, due to the development of experiment technologies and the deepening of theoretical studies, more and
more attentions have been focused on complex systems which involve long-range interactions \cite{lahaye,lauer1,trefzger,naini,schauss1}, multi-component gases \cite{altman1,soltan-panahi1}, frustrations \cite{eckardt-2010,pielawa,ye-2012}, and superlattice structures \cite{piel1,sebby-strabley1,folling1,cheinet1,jo1}. In such complex systems, there are some novel phases \cite{Lewenstein1}, including charge-density-wave (CDW) phase, supersolid phase \cite{SS_RMP}, paired superfluid phase, super-counter-fluid phase \cite{SCF} and so on. Phase diagrams of these complex systems are expected to be rather diverse and complex.

In order to study such rich phases in various optical superlattices systems, people have developed some numerical methods including QMC \cite{QMC_ED}, exact diagonalization \cite{QMC_ED} as well as density-matrix renormalization group method \cite{DMRG,FDMRG}, and analytical methods including decoupled mean-field theory \cite{DMF1,DMF2,DMF3,DMF4}, multisite mean-field theory \cite{MMF1,MMF2,MMF3,MMF4}, the cell strong-coupling expansion \cite{SCE1,SCE2}, the generalized Green¡¯s function method (GGFM)\cite{GGF1,GGF2,wei} as well as the GEPLT \cite{wang,wei}. Compared with the unbiased QMC results, it is found that the mean-filed theory underestimates the locations of the phase boundaries whereas the strong-coupling expansion method overestimates it.
Based on the same perturbation treatment, both GGFM and GEPLT have the potential to obtain the more accurate locations of the boundaries of the second-order phase transition, but from a technical standpoint, the GEPLT is easy to obtain higher-orders hopping corrections \cite{wang,wei}. However, the present GEPLT treatments \cite{wang,wei} have some technical issues that hinder its full application. These issues are required to be fixed to demonstrate the power of GEPLT framework.

Although the phase boundaries of ultracold Bose gases in bipartite superlattices have been analytically calculated up to second-order hopping \cite{wang} and have also been numerically obtained \cite{wei} up to higher-orders hopping via the process-chain approach with GEPLT. There still exist some subtle problems in theirs papers which need to be clarified.  In Wang et al.'s work \cite{wang}, the subtle problem is that when $\Delta \mu < 0.35U$ ($\Delta \mu$ is the difference of chemical potential for different sublattices), the full lobe of the MI phase (second-order result) can not be obtained. In Wei et al.'s work \cite{wei}, there are two subtle problems. One is that there is no odd-order corrections for the critical $t$. The other is that when $\Delta \mu =0$, the systems go back to the homogeneous systems, but the critical value of $t$ (see Eq.~10 in Wei et al.'s work \cite{wei}) for bipartite systems is not the same with the results (see Eq.~\ref{homo-boundary}) of the homogeneous systems. These subtle issues call for a more consistent and complete GEPLT treatment that includes high order corrections.

In this paper, we will clarify these technical issues (see Sec.~\ref{subtle}) and then propose a consistent and complete GEPLT to study the quantum phase transitions of ultracold Bose gases in bipartite superlattices. Using the corrected GEPLT,  we get the quantum phase boundaries up to third-order hopping for ultracold Bose gases in square and cubic superlattices. Our analytical results are in good agreement with the QMC simulations and the relative deviation of our third-order results from the QMC results is less than $9\%$ for square superlattices and $4\%$ for cubic superlattices.

\section{The model and Generalized Effective-Potential Landau Theory}\label{subtle}
 \begin{figure}[h!]
\centering
\includegraphics[width=0.7\linewidth]{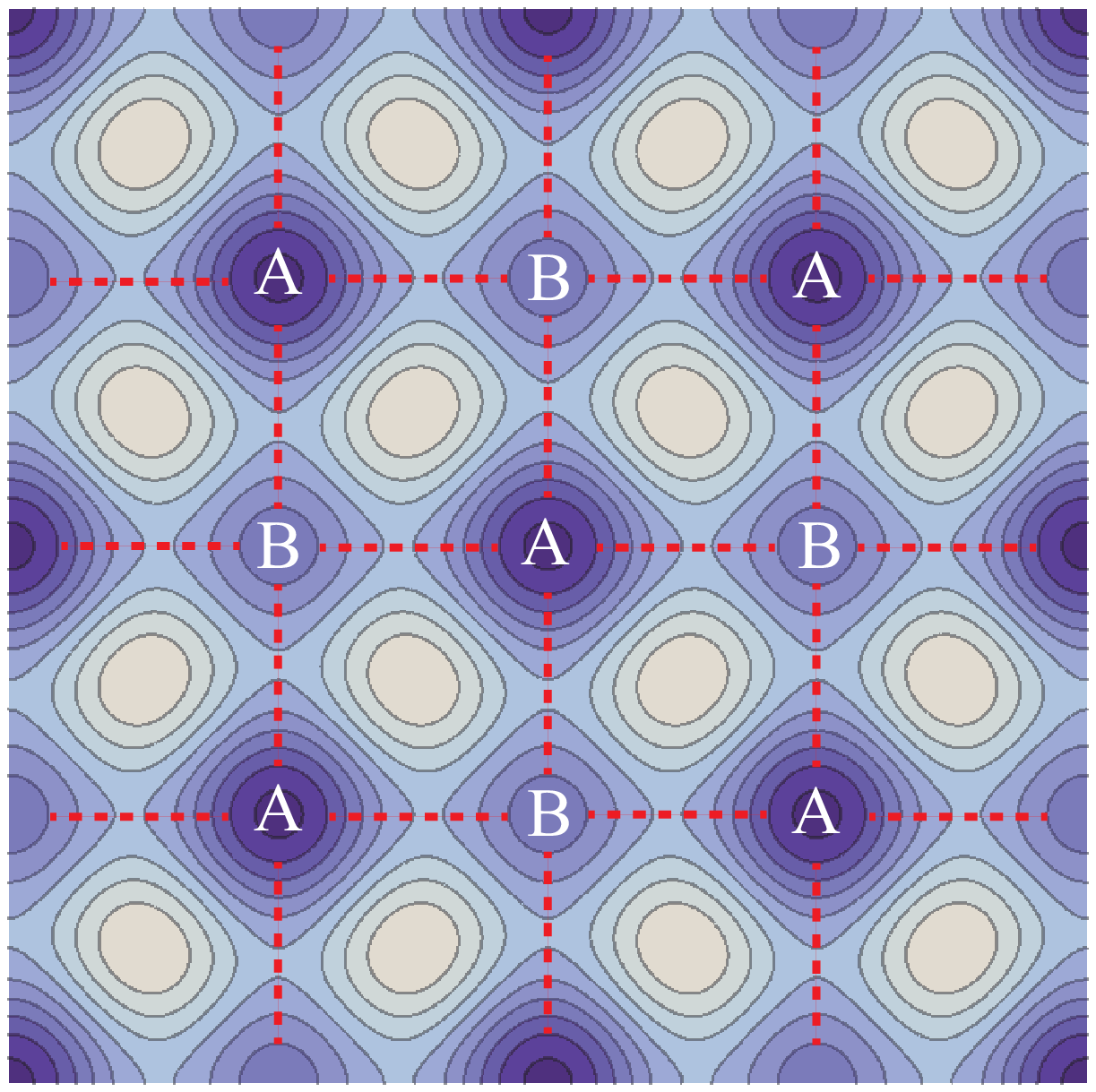}
\caption{The sketch of the square superlattice, the lattice sites A are deeper $\Delta \mu$ than lattice sites B.}
\label{superlattice}
\end{figure}
The systems with ultracold spinless bosonic atoms tapped in bipartite superlattices can be described by \cite{DMF1,wang}
\begin{eqnarray}
\hat{H}&=&-t\sum_{i\in A, j\in B}\left(\hat{a}^{\dag}_{i}\hat{a}_{j}+\hat{a}^{\dag}_{j}\hat{a}_{i}\right)+
\frac{U}{2}\sum_{i\in A,B}\hat{n}_{i}\left(\hat{n}_{i}-1\right)\nonumber \\
&& -\left(\mu+\Delta \mu\right)\sum_{i\in A}\hat{n}_{i}-\mu\sum_{i\in B}\hat{n}_{i}, \label{H}
\end{eqnarray}
where $t$ is the nearest-neighbor hopping amplitude,  $U$ denotes the
on-site repulsion, $\hat{a}^{\dag}_{i}$ is the creation operator and $\Delta \mu$ is the difference of chemical potential for sublattice $A$ and $B$.
The square superlattices  can be realized by choosing the trapping potential as \cite{SLP,wei}
\begin{eqnarray}
V(\mathbf{r})&=&-V_{0}\left[\cos^{2}\left(\frac{2\pi x}{\lambda}\right)+\cos^{2}\left(\frac{2\pi y}{\lambda}\right)\right. \nonumber \\
&&+2\cos\theta\cos\left(\frac{2\pi x}{\lambda}\right)\cos\left(\frac{2\pi y}{\lambda}\right)\left.\right],
\end{eqnarray}
where $\lambda$ is the wavelength of laser, and $\theta$ ($0<\cos\theta<1$) is the phase difference between the counterpropagating laser beams. The corresponding sketch of square superlattice is shown in Fig~\ref{superlattice}. The value of $\Delta \mu $ can be adjusted by changing the value of $\theta$. Hereinafter, without loss of generality, we set $\Delta \mu>0$ .

Let us first investigate the diagonal part of the system, whose Hamiltonian $\hat{H}_{0}$ reads
\begin{equation}
\hat{H}_{0}=\frac{U}{2}\!\!\!\sum_{i\in A,B}\!\!\hat{n}_{i}\left(\hat{n}_{i}-1\right)-\left(\mu+\! \Delta \mu\right)\!\sum_{i\in A}\hat{n}_{i}-\mu\!\sum_{i\in B}\hat{n}_{i}.
\end{equation}
The eigenvalue of $\hat{H}_{0}$ is $E_{n_{A},n_{B}}$  and the corresponding eigenstate is $\left|n_{A},n_{B}\right\rangle$,
where $n_{A(B)}$ denotes the filling factor of the sublattice A(B).  The eigen equation and $E_{n_{A},n_{B}}$ take the form
\begin{eqnarray}
 \hat{H}_{0}\left|n_{A},n_{n_{B}}\right\rangle\!\!\!&&= N_{s} E_{n_{A},n_{B}}\left|n_{A},n_{n_{B}}\right\rangle,\nonumber \\
 E_{n_{A},n_{B}}\!\!\!&&=\frac{U}{2}n_{A}(n_{A}-1)+\frac{U}{2} n_{B}(n_{B}-1)\nonumber \\
&&\quad-(\mu+\Delta\mu)n_{A}-\mu n_{B}.
\end{eqnarray}
The total number of lattice sites is $2 N_{s}$ and $N_{s}$ is the number of super unit cell. The ground states can be revealed via the game of throwing bosons into the superlattice. Bose atoms favor to stay in sublattice $A$ than sublattice $B$ for positive $\Delta\mu$. The first $N_{s}$ particles will fill up  sublattice A, and then the second $N_{s}$ bosons will fill up sublattice A or B that is determined by whether $E_{2,0}$ is larger than $E_{1,1}$ or not.
The state $\left|n,n\right\rangle$ is called MI state and the state $\left|n_{A},n_{B}\right\rangle$ with $n_{A}\neq n_{B}$ is called CDW state. In the case of  $\Delta\mu \in[0,U)$, the second $N_{s}$ bosons will fill up sublattice B, and the third $N_{s}$ bosons will go to sublattice A and so on. Thus it is evident that $n_{A}-n_{B}$ is always $1$ or $0$ in this case. In the case of $\Delta\mu>U$, there is no MI state, because in the $\rm{MI}(n,n)$ state the chemical potential needs to satisfy the constraint $U(n-1)\leq \mu\leq Un-\Delta \mu$ obtained by solving the inequations $E_{n+1,n}\geq E_{n,n} \leq E_{n,n-1}$.  This constraint is unsatisfied  when $\Delta\mu>U$, thus there is only CDW state. Generally speaking, if $U\left(n-1\right)\leq \Delta\mu \leq U n$ with $n\geq 2 $, the $n_{A}-n_{B}$ satisfy the constraint $ 1\leq n_{A}-n_{B}\leq n$.

In the case of $\Delta \mu <U$ or $\Delta \mu>U$,  the systems belong to different types, i.e., existence or nonexistence of the MI state.
In the next section, we will study the phase boundaries of bipartite superlattices with three representative ponits $\Delta \mu=0.1U,\, 0.5U,\, 1.5U$ to reveal this difference. In Wang et al.'s work \cite{wang}, they  have presented the QMC result for $\Delta \mu= 0.5U$, thus we can calculate the phase diagrams at this point to check the corrected GEPLT. The reason for choosing the case of $\Delta \mu = 0.1U$ is that in Wang et al.'s framework \cite{wang}, they declare that when $\Delta \mu < 0.35U$, the full lobe of the second-order corrected phase boundary of the MI phase  can not be obtained by GEPLT. The reason why Wang et al. failed to obtain this full MI lobe is that they make a tiny mistake for acquiring phase boundaries equations. It is very subtle in obtaining the phase boundaries equations by using GEPLT. We will clarify this  problem at below.

Before clarifying this subtle problem, we would like to give a short review of GEPLT introduced by Wang et al \cite{wang}. We add two components external  current vector $\mathbf{J}=\left(J_{A},J_{B}\right)^{T}$ into the above bipartite superlattice Bose-Hubbard Hamiltonian as follow:
\begin{eqnarray}
\hat{H}(\mathbf{J},\mathbf{J}^{\dag})&=&\hat{H}+\sum_{i\in A}\left(J_{A}\hat{a}^{\dag}_{i}+J_{A}^{\ast}\hat{a}_{i}\right)\nonumber \\
&&+\sum_{i\in B}\left(J_{B}\hat{a}^{\dag}_{i}+J_{B}^{\ast}\hat{a}_{i}\right). \label{HJJ}
\end{eqnarray}
By treating the hopping amplitude $t$, the external current $\mathbf{J}$ and $\mathbf{J}^{\dag}$ as  perturbations, the free energy of systems with external current can be written as
\begin{eqnarray}
F(\mathbf{J},\mathbf{J}^{\dag},t)&=&N_{s}\left(F_{0}(t)+\mathbf{J}^{\dag}C_{2}(t)\mathbf{J} \right. \nonumber \\
&&\left.+\mathbf{J}^{\dag}\mathbf{J}^{\dag}C_{4}(t)\mathbf{J}\mathbf{J}+\cdots\right)
\end{eqnarray}
at zero temperature, where the expansion coefficient $C_{2}(t)$ is a  second-order tensor which reads
\begin{eqnarray}
C_{2}(t)&=&\!\!\!\left(
           \begin{array}{cc}
             c_{2AA} & c_{2AB} \\
             c_{2BA} & c_{2BB} \\
           \end{array}
         \right) \nonumber \\
&=&\!\!\!\sum_{n=0}^{\infty}\!\left(\!
           \begin{array}{cc}
             \alpha_{2AA}^{(2n)}(-t)^{2n} & \alpha_{2AB}^{(2n+1)}(-t)^{2n+1} \\
             \alpha_{2BA}^{(2n+1)}(-t)^{2n+1} & \alpha_{2BB}^{(2n)}(-t)^{2n} \\
           \end{array}
         \!\right)\!.
\end{eqnarray}
 It is obvious that $c_{2AB}$ is equal to $c_{2BA}$, for they describe Hermitian processes. Because the ground state of non-perturbation part is MI or CDW state, the free energy $F(\mathbf{J},\mathbf{J}^{\dag},t)$ is only constituted by pairing of the external current $\mathbf{J}$ and $\mathbf{J}^{\dag}$.

 \begin{figure}[h!]
\centering
\includegraphics[width=0.8\linewidth]{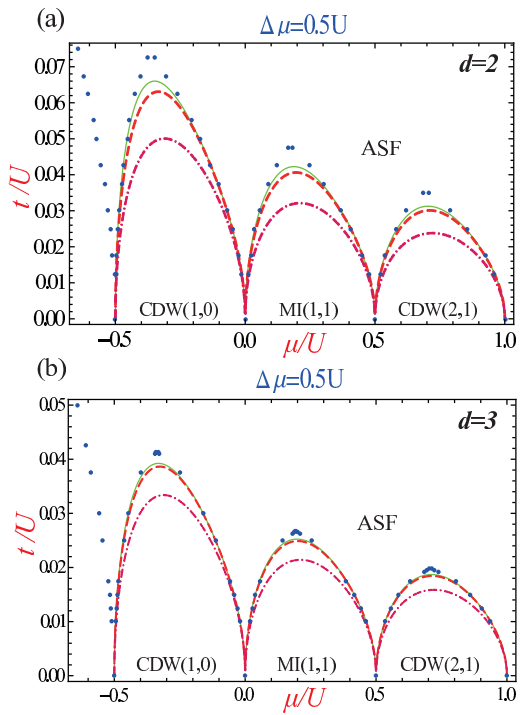}
\caption{(Color online)The phase diagram of ultracold Bose gases in square ($d=2$) (a), cubic ($d=3$)  (b) bipartite superlattice  for the case of  $\Delta\mu=0.5U$.
 The green solid  line is the third-order calculation, the red dashed
line is the second-order result, and the pink dot-dashed line is the
first-order (mean-field) result. The blue dots line is the QMC result\cite{wang}. }.
\label{phase-diagram-superlattice1}
\end{figure}

Due to the bipartite lattice structure of systems, the order parameter can be defined as two components vector $\Psi=(\psi_{A},\psi_{B})$, where $\psi_{A}=\langle \hat{a}_{i}\rangle$ for $i\in A$ and  $\psi_{B}=\langle \hat{a}_{i}\rangle$ for $i\in B$. It is easy to find that order parameter reads $\psi_{A/B}=\frac{1}{N_{s}}\frac{\partial F(\mathbf{J},\mathbf{J}^{\dag},t)}{\partial J^{\ast}_{A/B}}$. After Legendre transformation,
we have the  effective potential
\begin{equation}
\Gamma (\Psi,\Psi^{\dag},t)=\frac{1}{N_{s}}F(\mathbf{J},\mathbf{J}^{\dag},t)-\Psi \mathbf{J}^{\dag} - \Psi^{\dag} \mathbf{J}.
\end{equation}
composed of free energy $F(\mathbf{J},\mathbf{J}^{\dag},t)$.
Thus, the external current can be calculated from the effective potential $\Gamma (\Psi,\Psi^{\dag},t)$  by $J_{A/B}=\frac{\partial \Gamma (\Psi,\Psi^{\dag},t)}{\partial \psi^{\ast}_{A/B}}$. Moreover, the effective potential $\Gamma (\Psi,\Psi^{\dag},t)$  can be rewritten as a function of $\Psi$ and $\Psi^{\dag}$
\begin{equation}
\Gamma (\Psi,\Psi^{\dag},t)\!=\!F_{0}(t)\!+\Psi^{\dag} \rm{A}_{2}(t) \Psi \!+ \Psi^{\dag} \Psi^{\dag} \rm{A}_{4}(t) \Psi \Psi+\cdots,
\end{equation}
where the expansion coefficient matrix $\rm{A}_{2}(t)$ is the inverse of $- C_{2}(t)$ \cite{wang}. According to Landau theory, the phase boundaries of this systems can be obtained by setting
\begin{equation}
\det \rm{A}_2(t) = \frac{-1}{c_{2AA}c_{2BB}-c_{2AB}c_{2BA}}=0.
\end{equation}
Hence, the critical value of $t$ can be determined by the radius of convergence of series $c_{2AA}c_{2BB}-c_{2AB}c_{2BA}$.

Wang et al. make an easy mistake on obtaining the phase boundaries via $\det \rm{A}_{2}(t)=0$. In the case of $\Delta \mu=0$, the phase boundaries of superlattice systems calculated by Wang et al., go back to the phase boundaries of the homogeneous systems which have been calculated by  F. E. A. dos Santos et al. via effective-potential Landau theory \cite{santos}. Unfortunately, the results obtained by F. E. A. dos Santos et al. are  incorrect and thus Wang et al. make the same mistake with F. E. A. dos Santos et al. The arguments for F. E. A. dos Santos et al.'s mistake are presented as below.

F. E. A. dos Santos et al., present a powerful method \cite{santos}(effective-potential Landau theory) to determine the location of seconde-order phase transition boundaries for high-dimensional single-component bose systems, such as square and cubic lattice as well as triangular, hexagonal and kagom\'{e} lattice \cite{lin1}, but they make an easy mistake on the details for obtaining the phase boundaries equations. In F. E. A. dos Santos et al.'s work ~\cite{santos}, if the equation $\frac{1}{c_{2}(t)}=(\alpha^{(0)}_{2})^{-1}(1+x+x^{2}+\cdots)$ (see Ref.~\cite{santos} ) is tenable that implicates $x\ll 1$, where $x=(\alpha^{(0)}_{2})^{-1}(\alpha^{(1)}_{2} t- \alpha^{(2)}_{2} t^{2}+\alpha^{(3)}_{2} t^{3}+\cdots)$ and $c_{2}(t)=\sum_{n}^{\infty}\alpha^{(n)}_{2}(-t)^{n}=\alpha^{(0)}_{2}\left(1-x\right)$. The $c_{2}(t)$ is the coefficient of the free energy with a pair current $|J|^{2}$ \cite{santos}, which includes all-order hopping corrections.  According to Landau theory, the phase boundaries can be obtained by solving the equation $1/c_{2}(t)=0$ \cite{santos} that  means the $c_{2}(t)\rightarrow \infty $ on the phase boundaries. Obviously, on the phase boundaries, $x$ will be diverging and this property is inconsistent with $x\ll 1$. So, it's not the right way (solving the equation $1/c_{2}(t)=0$ by the series expanding $1/c_{2}(t)$) to obtain the phase boundaries in  F. E. A. dos Santos et al.'s work~\cite{santos}. The correct phase boundaries \cite{Teichmann1,Teichmann2,lin1} is given by
\begin{equation}
t_{c}^{(n)}=-\frac{\alpha^{(n-1)}_{2}}{\alpha^{(n)}_{2}}, \label{homo-boundary}
\end{equation}
which is the radius of convergence of $c_{2}(t)$ determined by the d'Alembert's ratio test.

In the above-mentioned arguments, we have clarified the minor mistake made by F. E. A. dos Santos et al. and Wang et al. to determine the phase boundaries of the homogeneous systems and bipartite superlattices systems respectively. At below, we will show the right way to handle this subtle problem.
Up to the fourth-order hopping, the series $c_{2AA}c_{2BB}\!-c_{2AB}c_{2BA}$ reads
\begin{eqnarray}
&&c_{2AA}c_{2BB}\!-c_{2AB}c_{2BA}\nonumber \\
&=&\bigg[\!a+\alpha^{(1)}_{2AB}t+bt^{2}+\alpha^{(3)}_{2AB}t^{3}+ct^{4}+\cdots\!\bigg] \nonumber \\
&&\times\!\bigg[\!a-\!\alpha^{(1)}_{2AB}t+bt^{2}-\alpha^{(3)}_{2AB}t^{3}+ct^{4}-\cdots\bigg],
\end{eqnarray}
where the coefficient $a$, $b$ and $c$ read
\begin{eqnarray}
&&\!\!\!\!\!\!a=\left(\alpha_{2AA}^{(0)}\alpha_{2BB}^{(0)}\right)^{1/2},
b=\frac{\alpha_{2AA}^{(0)}\alpha_{2BB}^{(2)}\!+\alpha_{2AA}^{(2)}\alpha_{2BB}^{(0)}}{2a}, \nonumber \\
&&\!\!\!\!\!\!c=\frac{\alpha_{2AA}^{(0)}\alpha_{2BB}^{(4)}\!+\alpha_{2AA}^{(4)}\alpha_{2BB}^{(0)}+\alpha_{2AA}^{(2)}\alpha_{2BB}^{(2)}-b^{2}}{2a}.
\end{eqnarray}
Since $\left(\alpha_{2AA}^{(0)}\alpha_{2BB}^{(0)}\right)^{1/2}$, $\alpha^{(1)}_{2AB}$, $\alpha_{2AA}^{(0)}\alpha_{2BB}^{(2)}\!+\alpha_{2AA}^{(2)}\alpha_{2BB}^{(0)}$ and $\alpha^{(3)}_{2AB}$ are all positive,  this series can be rewritten as
 $c_{2AA}c_{2BB}-c_{2AB}c_{2BA}=\left(\sum_{n=0}^{\infty}a_{n}t^{n}\right)\left(\sum_{n=0}^{\infty}|a_{n}|t^{n}\right)$, where $\sum_{n=0}^{\infty}a_{n}t^{n}$ reads
\begin{eqnarray}
\sum_{n=0}^{\infty}a_{n}t^{n}=\left[\!a-\alpha^{(1)}_{2AB}t+bt^{2}\!\!-\alpha^{(3)}_{2AB}t^{3}\!+ct^{4}\cdots\right].
\end{eqnarray}
Importantly, the series $\sum_{n=0}^{\infty}|a_{n}|t^{n}$ include all-order hopping processes that is very obvious when $\Delta \mu =0$ (when $\Delta \mu =0$, $\sum_{n=0}^{\infty}|a_{n}|t^{n}=-\sum_{n=0}^{\infty}\alpha^{(n)}_{2}(-t)^{n}$, here $\alpha^{(0)}_{2AA}<0$ and $\alpha^{(0)}_{2BB}<0$ ).
Moreover, Power series $\sum_{n=0}^{\infty}a_{n}t^{n}$ and $\sum_{n=0}^{\infty}|a_{n}|t^{n}$ have the same radius of convergence $r=\lim_{n\rightarrow \infty}|a_{n}|/|a_{n+1}|$ determined by d'Alembert's ratio test.  Thus, the radius of convergence of $c_{2AA}c_{2BB}-c_{2AB}c_{2BA}$ is only determined by the series $\sum_{n=0}^{\infty}|a_{n}|t^{n}$.
\begin{figure}[h!]
\centering
\includegraphics[width=0.8\linewidth]{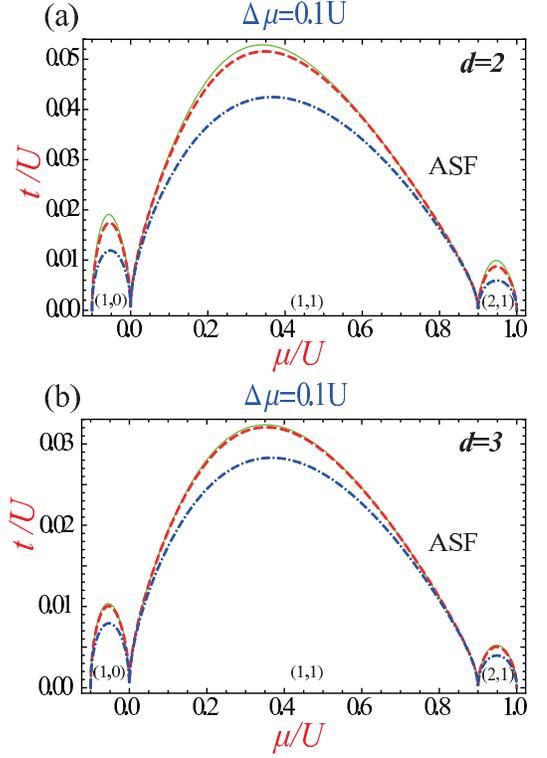}
\caption{(Color online)The phase diagram of ultracold Bose gases in square (a), cubic (b) superlattice  for the case of  $\Delta\mu=0.1U$.
 The green solid line is the third-order calculation, the red dashed
line is the second-order result, and the blue dot-dashed line is the
first-order (mean-field) result. }.
\label{phase-diagram-superlattice2}
\end{figure}

On the other hand, it is not suitable to fix the radius of convergence by directly handling
$c_{2AA}c_{2BB}\!-c_{2AB}c_{2BA}=\sum_{n=0}^{\infty} d_{n}(t^2)^{n}$, which include only even-order term of $t$. The coefficient $d_{n} (t^2)^{n}$ fails to describe the 2n-$\rm{th}$-order hopping process. To see this, let us consider $\Delta \mu =0$. It is clear that  $d_{2}=2\alpha^{(0)}_{2}\alpha^{(2)}_{2}-(\alpha^{(1)}_{2})^{2}$ does not correspond to second-order hopping process. Thus, we can not directly calculate the  radius of convergence by $r_{d}=\lim_{n\rightarrow \infty}|d_{n}|/|d_{n+1}|$.  Because Wei et al. \cite{wei} use the wrong radius of convergence $r_{d}$, there are two physical problems in Wei et al.'s \cite{wei} work. One is that there is no odd-order correction terms of critical $t$.  The other is that if we set $\Delta \mu =0$, the systems go back to the homogeneous systems, but the phase boundaries equations of bipartite systems (see Eq.~10 in Wei et al.'s work \cite{wei}) is not same with the phase boundaries equations of the homogeneous systems. Based on d'Alembert's ratio test,
the correct phase boundaries of bipartite superlattice systems are given by
\begin{equation}
t^{(1)}_{c}=\frac{\sqrt{\alpha_{2AA}^{(0)}\alpha_{2BB}^{(0)}}}{\alpha_{2AB}^{(1)}},\label{mean-field}
\end{equation}
\begin{equation}
t^{(2)}_{c}=\frac{2\sqrt{\alpha_{2AA}^{(0)}\alpha_{2BB}^{(0)}}\alpha_{2AB}^{(1)}}{\alpha_{2AA}^{(0)}\alpha_{2BB}^{(2)}+\alpha_{2BB}^{(0)}\alpha_{2AA}^{(2)}},
\label{2-order}
\end{equation}
\begin{equation}
t^{(3)}_{c}=\frac{\alpha_{2AA}^{(0)}\alpha_{2BB}^{(2)}+\alpha_{2BB}^{(0)}\alpha_{2AA}^{(2)}}{2\sqrt{\alpha_{2AA}^{(0)}\alpha_{2BB}^{(0)}}\alpha_{2AB}^{(3)}},
\label{3-order}
\end{equation}
which are the radius of convergence of $\sum_{n=0}^{\infty}|a_{n}|t^{n}$. Here $t^{(i)}_{c}$ is the n-$\rm{th}$ order phase boundaries and the  first-order result is the same as the mean-field result \cite{Iskin,GGF1} (By setting the nearest-neighbor interaction $V=0$ and $\mu_{A}=\mu+\Delta\mu$ in Refs.~\cite{Iskin,GGF1} ). When $\Delta \mu =0$, the Eqs.~\ref{mean-field}, \ref{2-order} and \ref{3-order} of phase boundaries  are consistent with  the Eq.~\ref{homo-boundary} of phase boundaries of the homogeneous systems \cite{Teichmann1,Teichmann2,lin1}. It is an independent check that proves our results.

\section{quantum phase diagram }
The perturbative coefficients $\alpha^{(n)}_{2ij}$ ($i,j \in A,B$) can be calculated by the Raylieigh-Schrodinger perturbation
expansion via diagrammatic representation, which have been gracefully and detailedly introduced \cite{santos,lin1,wang}. Thus, we do not give the detail for calculating these coefficients $\alpha^{(n)}_{2ij}$, but we would like to present the first four orders of diagrammatic expressions of $\alpha^{(n)}_{2ij}$, which are shown in the Table.~\ref{table-1}.

\begin{table}[h!]
\caption{The diagrams of the coefficients $\alpha^{(n)}_{2ij}$ for bipartite sublattice systems, where $Z$ is the coordination number of the systems.}%
\begin{tabular}{|c|c|c|c|}
  \hline
\raisebox{-0.1cm}{$\alpha_{2AA}^{(0)}$} & $\raisebox{-0.22cm}{\includegraphics[width=1.8cm]{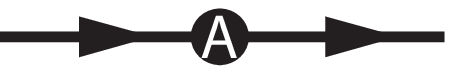}}$& \raisebox{-0.1cm}{$\alpha_{2BB}^{(0)}$}
& $\raisebox{-0.22cm}{\includegraphics[width=1.8cm]{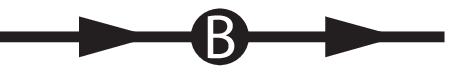}}$ \\\hline
\raisebox{-0.08cm}{$\alpha_{2AB}^{(1)}$} & $\raisebox{-0.15cm}{Z}\raisebox{-0.22cm}{\includegraphics[width=2.6cm]{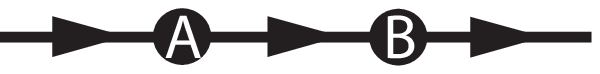}}$& \raisebox{-0.08cm}{$\alpha_{2BA}^{(1)}$}& $\raisebox{-0.15cm}{Z}\raisebox{-0.22cm}{\includegraphics[width=2.6cm]{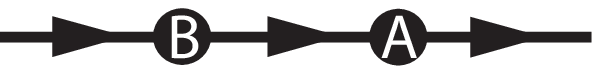}}$ \\\hline
\raisebox{0.2cm}{$\alpha_{2AA}^{(2)}$} & \multicolumn{3}{c|}{$\raisebox{0.1cm}{Z(Z-1)} \raisebox{-0.01cm}{\includegraphics[width=3.6cm]{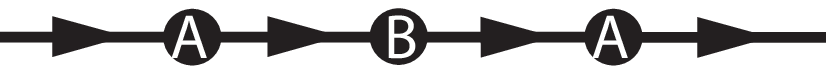}}
\raisebox{0.1cm}{+Z}\raisebox{-0.01cm}{\includegraphics[width=1.6cm]{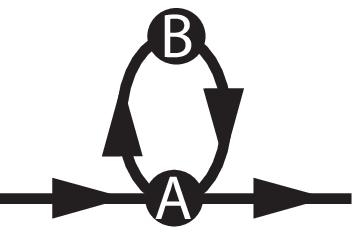}}$}\\\hline
 \raisebox{0.2cm}{$\alpha_{2BB}^{(2)}$} & \multicolumn{3}{c|}{$\raisebox{0.1cm}{Z(Z-1)}
 \raisebox{-0.01cm}{\includegraphics[width=3.6cm]{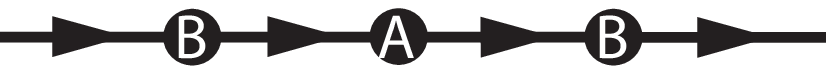}}\raisebox{0.1cm}{+Z}\raisebox{-0.01cm}{\includegraphics[width=1.6cm]{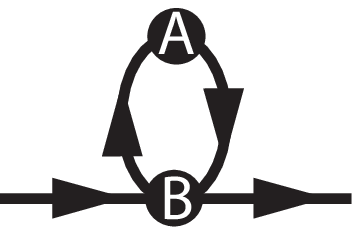}}$} \\ \hline
  & \multicolumn{3}{c|}{\raisebox{-0.25cm}{Z(Z-1)(Z-1)}\raisebox{-1.2cm}{\includegraphics[width=4.5cm]{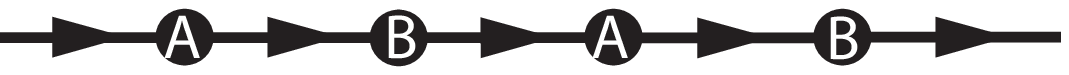}}} \\
\raisebox{0.9cm}{$\alpha_{2AB}^{(3)}$} & \multicolumn{3}{c|}{\raisebox{0.15cm}{+Z(Z-1)}\raisebox{0.03cm}{\includegraphics[width=2.3cm]{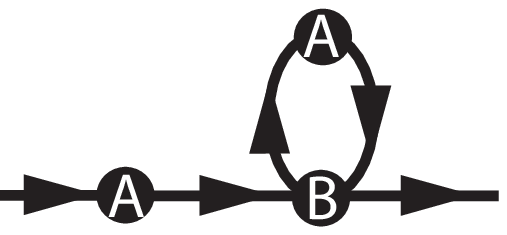}}
    \raisebox{0.15cm}{+Z(Z-1)}\raisebox{0.03cm}{\includegraphics[width=2.3cm]{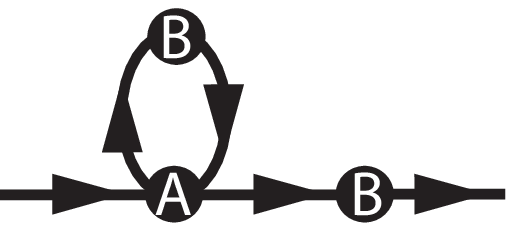}}}\\
 &\multicolumn{3}{c|}{\raisebox{-0.25cm}{+Z}\raisebox{-0.6cm}{\includegraphics[width=3.7cm]{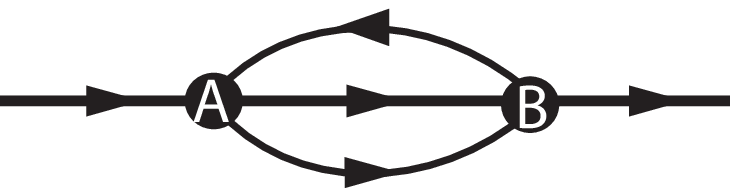}}}
  \\ \hline
    & \multicolumn{3}{c|}{\raisebox{-0.25cm}{Z(Z-1)(Z-1)}\raisebox{-1.2cm}{\includegraphics[width=4.5cm]{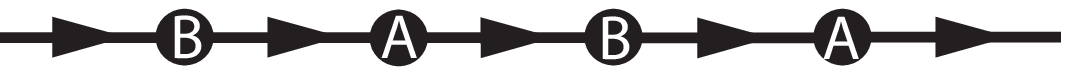}}} \\
\raisebox{0.9cm}{$\alpha_{2BA}^{(3)}$} & \multicolumn{3}{c|}{\raisebox{0.15cm}{+Z(Z-1)}\raisebox{0.03cm}{\includegraphics[width=2.3cm]{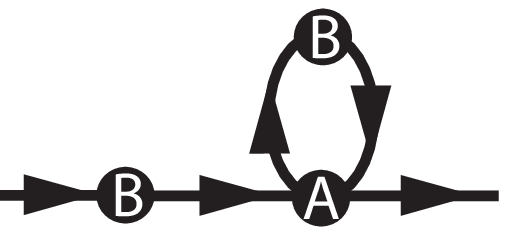}}
    \raisebox{0.15cm}{+Z(Z-1)}\raisebox{0.03cm}{\includegraphics[width=2.3cm]{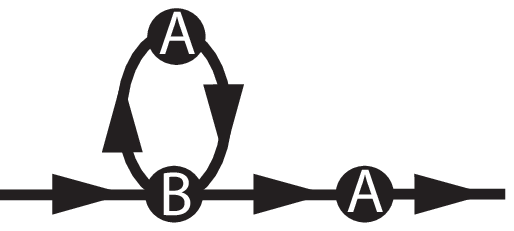}}}\\
 &\multicolumn{3}{c|}{\raisebox{-0.25cm}{+Z}\raisebox{-0.6cm}{\includegraphics[width=3.7cm]{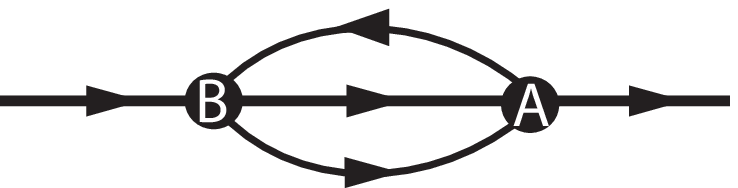}}} \\
  \hline
\end{tabular}
\label{table-1}
\end{table}

Before showing the phase diagrams of this systems, we would like to discuss the property of SF phase in bipartite superlattice. When $t\gg U$, the eigenstates of systems can be written as
$\left| \Psi \right\rangle \sim \prod_{i}\left[\exp{\{\sqrt{n_{i}}\hat{a}_{i}^{\dag}\}}|0\rangle_{i}\right]$\cite{RMP}, where $n_{i}$ denotes the condensate density on
the site $i$. Using this wave function, we can get free energy of each super unit cell, which takes the form
\begin{eqnarray}
E &\sim &-2t\sqrt{n_{A}}\sqrt{n_{B}}-(\mu+\Delta\mu)n_{A} \nonumber\\
&&+\frac{U}{2}n_{A}(n_{A}-1)-\mu n_{B} +\frac{U}{2}n_{B}(n_{B}-1)\nonumber\\
&\sim &\overline{n}(-2t-2\mu-\Delta\mu+U(\overline{n}-1))\nonumber\\
&&+\frac{t(\delta n)^{2}}{\overline{n}}-\Delta\mu \delta n +U(\delta n)^{2},\label{free-asf1}
\end{eqnarray}
where  $n_{A}=\overline{n}+\delta n$ and $n_{B}=\overline{n}-\delta n$ is the density on sublattice $A$ and $B$,  $\overline{n}$ is average density
and $\delta n$ is density fluctuation.  In the final step of Eq.~\ref{free-asf1}, we use the condition $\delta n/\overline{n}\ll 1$. From this free energy, we can find that the free energy have a minimum at  density fluctuation $\delta n =\frac{\Delta\mu }{2t/\overline{n} +2U}$.
\begin{figure}[h!]
\centering
\includegraphics[width=0.8\linewidth]{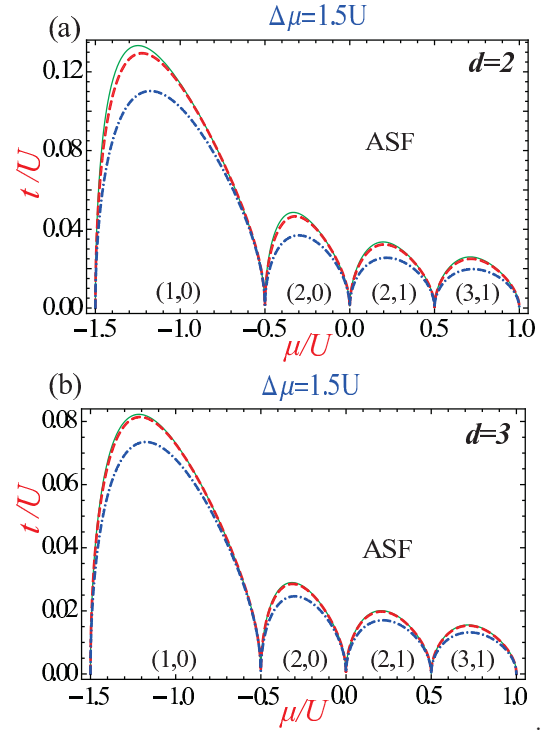}.
\caption{The phase diagram of ultracold Bose gases in square and cubic superlattices systems at $\Delta\mu=1.5U$. The green solid line is the third-order calculation,
the red dashed line is the second-order result, and the blue dot-dashed line is the first-order (mean-field) result }.
\label{phase-diagram-superlattice3}
\end{figure}
This means that homogeneous SF phase will not appear in this systems for any finite value of $\Delta\mu$ and there is only anisotropic superfluid (ASF).
In Wei et al.'s work \cite{wei}, there always are extra peaks at $\left(\pm\pi,\pm\pi\right)$ in time-of-flight pictures in SF phase for bipartite superlattice systems. These extra peaks reflect the inhomogeneity of the SF phase that is in accord with our argument.
Using  Eqs.(\ref{mean-field}), (\ref{2-order}) and (\ref{3-order}), we can get the first, second and third order phase boundaries of
 CDW(MI)-ASF quantum phase transition for ultracold Bose gases in square and cubic superlattice systems.
From Fig.~\ref{phase-diagram-superlattice1}, we easily find our analytical result accord with the QMC result \cite{wang} very well at $\Delta \mu =0.5U$.
The relative deviation of our third-order results from the QMC results is less than $9\%$ for square superlattices and $4\%$ for cubic superlattices. Furthermore, using these new phase boundaries Eqs.(\ref{mean-field}), (\ref{2-order}) and (\ref{3-order}), we can show the phase boundaries at $\Delta \mu =0.1U <0.35U$ (see Fig.~\ref{phase-diagram-superlattice2}), but it can not be obtained in  Wang et al.'s work \cite{wang}. Comparing Fig.~\ref{phase-diagram-superlattice1} and Fig.~\ref{phase-diagram-superlattice2}, it is clear that the region of $\rm{MI(1,1)}$ increases with decreasing $\Delta \mu$ while the region of $\rm{CDW}(1,0)$ or $\rm{CDW}(2,1)$ decreases correspondingly, which are as expected. In the case of $\Delta \mu>U$, there is no MI state in systems. Taking $\Delta \mu=1.5U$ as an example, we show its phase boundaries in Fig.~\ref{phase-diagram-superlattice3}.

Moreover, high order correction for different dimensional systems ($d=2,\,3$) are investigated. As shown in Figs.~\ref{phase-diagram-superlattice1}, \ref{phase-diagram-superlattice2} and \ref{phase-diagram-superlattice3},  higher order corrections are smaller in $d=3$ (cubic superlattice) than that in $d=2$ (square superlattice) systems. This phenomenon is due to the fact that the effect of quantum fluctuation is smaller in higher dimensionality.

\section{Summary}
In this paper, we clarify a tiny and easy mistake made by Wang et al.\cite{wang}  and Wei et al. \cite{wei} when they use GEPLT to study the phase boundaries of square and cubic superlattice systems respectively.  After clarifying these technical issues, GEPLT becomes more consistent and complete, valid for all value of $\Delta \mu$. Specially, the critical $t$ obtained by corrected GEPLT coincides with the result of homogeneous systems \cite{lin1} when $\Delta \mu =0$. Using the corrected phase boundaries equations, we have presented the quantum phase boundaries up to third-order hopping for ultracold Bose gases in square and cubic superlattices. Our analytical results are in excellent agreement with the QMC simulations and the relative deviation of our third-order results from the QMC results is less than $9\%$ for square superlattices and $4\%$ for cubic superlattices. Furthermore, we emphasize that homogeneous SF phase does not appear when $\Delta\mu \neq 0$ and  only ASF phase exists in  such superlattice systems.

\section*{Acknowledgement}
Z L acknowledges inspiring discussions with Yan Chen, Ying Jiang and also thanks Tao Wang for providing the QMC data and useful discussions. Z L wish also to thank  Dan Bo Zhang for reading and providing useful comments on this manuscript. This work was supported by National Natural Science Foundation of China [Grant Nos.~11074043, 11274069] and by the State Key Programs of China (Grant Nos.~2012CB921604 and 2009CB929204).


\begin{thebibliography}{99}
\bibitem{Lewenstein} M. Lewenstein, A. Sanpera, V. Ahufinger, B. Damski,
A. Sen(De), and U. Sen, Adv. Phys. \textbf{56}, 243 (2007).
\bibitem{RMP}I. Bloch, J. Dalibard, and W. Zwerger, Rev. Mod. Phys. \textbf{80}, 885 (2008).

\bibitem{synthetic-gauge-field1}J. Dalibard, F. Gerbier, G. Juzeliunas, and P. \"{o}hberg, Rev. Mod. Phys.  \textbf{83}, 1523 (2011).
\bibitem{SOC1}  V. Galitski, and I. B. Spielman,  Nature (London) \textbf{494}, 49 (2013).
\bibitem{synthetic-gauge-field2}  N. Goldman, G. Juzeliunas, P. \"{o} hberg, and I. B.  Spielman, Rep. Prog. Phys. \textbf{77}, 126401 (2014).
\bibitem{SOC2}  H. Zhai, Rep. Prog. Phys. \textbf{78}, 026001 (2015).
\bibitem{Eckardt} A. Eckardt, Rev. Mod. Phys. \textbf{89}, 011004 (2017).
\bibitem{QS1}I. Buluta, and F. Nori,  Science \textbf{326}, 108 (2009).
\bibitem{QS2}I. M. Georgescu, S. Ashhab, and F. Nori, Rev. Mod. Phys.  \textbf{86}, 153 (2014).
\bibitem{QS3}C. Gross and I. Bloch, Annu. Rev. Cold At. Mol. \textbf{3}, 181 (2015).

\bibitem{BHM1}M. P. A. Fisher, P. B. Weichman, G. Grinstein, and D. S. Fisher, Phys. Rev. B \textbf{40}, 546 (1989).
\bibitem{BHM2} D. Jaksch, C. Bruder, J. I. Cirac, C. W. Gardiner, and P. Zoller,  Phys. Rev. Lett. \textbf{81}, 3108 (1998).
\bibitem{Greiner} M. Greiner, O. Mandel, T. Esslinger, T. W.  H\"{a}nsch, and I. Bloch, Nature (London) \textbf{415}, 39 (2002).

\bibitem{lahaye} T. Lahaye, C. Menotti, L. Santos, M. Lewenstein, and T. Pfau, Rep. Prog. Phys. {\bf 72}, 126401 (2009).
\bibitem{trefzger} C. Trefzger, C. Menotti, B. Capogrosso-Sansone, and M. Lewenstein, J. Phys. B: At. Mol. Opt. Phys. {\bf 44}, 193001 (2011).
\bibitem{lauer1} A. Lauer, D. Muth, and M. Fleischhauer, New J. Phys. {\bf 14}, 095009 (2012).
\bibitem{schauss1}  P. Schau\ss, M. Cheneau, M. Endres, T. Fukuhara, S. Hild, A. Omran, T. Pohl, C. Gross, S. Kuhr, and I, Bloch, Nature (London) {\bf 491}, 87 (2012).
\bibitem{naini} A. Safavi-Naini, S. G. Soyler, G. Pupillo, H. R. Sadeghpour, and B. Capogrosso-Sansone, New J. Phys. {\bf 15}, 013036 (2013).
\bibitem{altman1} E. Altman, W. Hofstetter, E. Demler, and M. D. Lukin, New J. Phys.  {\bf 5}, 113 (2003).
\bibitem{soltan-panahi1} P. Soltan-Panahi, D. L\"uhmann, J. Struck, P. Windpassinger, and K. Sengstock, Nat. Phys. {\bf 8}, 71 (2012).
\bibitem{eckardt-2010}A. Eckardt, P. Hauke, P. Soltan-Panahi, C. Becker, K. Sengstock, and M. Lewenstein, EuroPhys. Lett. {\bf 89}, 10010 (2010).
\bibitem{pielawa} S. Pielawa, E. Berg, and S. Sachdev, Phys. Rev. B {\bf 86}, 184435 (2012).
\bibitem{ye-2012} J. Ye, K. Zhang, Y. Li, Y. Chen, and W. Zhang, Ann. Phys.  {\bf 328}, 103 (2013).
\bibitem{piel1}  S. Peil, J. V. Porto, B. Laburthe Tolra, J. M. Obrecht, B. E. King, M. Subbotin, S. L. Rolston, and W. D. Phillips, Phys. Rev. A {\bf 67}, 051603(R) (2003).
\bibitem{sebby-strabley1} J. Sebby-Strabley, M. Anderlini, P. S. Jessen, and J. V. Porto, Phys. Rev. A {\bf 73}, 033605 (2006).
\bibitem{folling1} S. F\"olling, S. Trotzky, P. Cheinet, M. Feld, R. Saers, A. Widera, T. M\"uller, and I. Bloch, Nature (London) {\bf 448}, 1029 (2007).
\bibitem{cheinet1} P. Cheinet, S. Trotzky, M. Feld, U. Schnorrberger, M. Moreno-Cardoner, S. F\"olling, and I. Bloch, Phys. Rev. Lett. {\bf 101}, 090404 (2008).
\bibitem{jo1}G. B. Jo, J. Guzman, C. K. Thomas, P. Hosur, A. Vishwanath, and D. M. Stamper-Kurn, Phys. Rev. Lett. {\bf 108}, 045305 (2012).

\bibitem{Lewenstein1}O. Dutta, M. Gajda, P. Hauke, M. Lewenstein,Dirk-S\"{o}ren L\"{u}hmann, B. A.  Malomed, T. Sowinski,
and J. Zakrzewski, Rep. Rrog. Phys. \textbf{78}, 066001 (2015).
\bibitem{SS_RMP} M. Boninsegni, and N. V. Prokof'ev, Rev. Mod. Phys. \textbf{84}, 759 (2012).
\bibitem{SCF} A. B. Kuklov, and B. V. Svistunov,  Phys. Rev. Lett. {\bf 90}, 100401 (2003).


\bibitem{QMC_ED}V. G. Rousseau, D. P. Arovas, M. Rigol, F. Hebert, G. G. Batrouni, and R. T. Scalettar, Phys. Rev. B \textbf{73}, 174516 (2006).
\bibitem{DMRG}G. Roux, T. Barthel, I. P. McCulloch, C. Kollath, U. Schollw\"{o}ck,  and T. Giamarchi, Phys. Rev. A \textbf{78}, 023628 (2008).
\bibitem{FDMRG}A. Dhar, T. Mishra, R. V. Pai, and B. P. Das, Phys. Rev. A \textbf{83}, 053621 (2011).

\bibitem{DMF1}P. Buonsante,  and A. Vezzani, Phys. Rev. A \textbf{70}, 033608 (2004).
\bibitem{DMF2}J. M. Hou, Mod. Phys. Lett. B \textbf{23}, 25 (2009).
\bibitem{DMF3}B. L. Chen, S. P. Kou, Y. Zhang, and S. Chen, Phys. Rev. A \textbf{81}, 053608 (2010).
\bibitem{DMF4}A. Dhar,  M. Singh, R. V. Pai, and B. P. Das, Phys. Rev. A \textbf{84}, 033631 (2011).


\bibitem{MMF1}P. Buonsante, V. Penna, and A. Vezzani, Laser Phys. \textbf{15}, 361 (2005).
\bibitem{MMF2}D. Muth, A. Mering, and M. Fleischhauer, Phys. Rev. A \textbf{77}, 043618 (2008).
\bibitem{MMF3}P. Pisarski, R. M. Jones, and R. J. Gooding, Phys. Rev. A \textbf{83}, 053608 (2011).
\bibitem{MMF4}T. McIntosh, P. Pisarski, R. J. Gooding and E. Zaremba,  Phys. Rev. A \textbf{86}, 013623 (2012).

\bibitem{SCE1}P. Buonsante, and A. Vezzani, Phys. Rev. A  \textbf{72}, 013614 (2005).
\bibitem{SCE2}P. Buonsante, V. Penna, and A. Vezzani, Phys. Rev. A \textbf{72}, 031602(R) (2005).

\bibitem{GGF1}Z. Lin, J. Zhang, and Y. Jiang, Front. Phys. \textbf{13}, 136401 (2018).
\bibitem{GGF2}J. Zhang, and Y. Jiang,  Laser Phys. \textbf{26}, 095501 (2016).
\bibitem{wei}Wei. F, J. Zhang, and Y. Jiang, Europhys. Lett. \textbf{113}, 16004 (2016).

\bibitem{wang} T. Wang, X. F. Zhang, S. Eggert, and A. Pelster, Phys. Rev. A \textbf{87}, 063615 (2013).
\bibitem{lin1}Z. Lin, J. Zhang, and Y. Jiang, Phys. Rev. A \textbf{85}, 023619 (2012).
\bibitem{SLP}S. Paul, and E. Tiesinga, Phys. Rev. A  \textbf{88}, 033615 (2013).
\bibitem{santos}F. E. A. dos Santos, and A. Pelster, Phys. Rev. A \textbf{79}, 013614 (2009).
\bibitem{Teichmann1}N. Teichmann, D. Hinrichs, M. Holthaus, and A. Eckardt, Phys. Rev. B \textbf{79}, 224515 (2009).
\bibitem{Teichmann2}N. Teichmann, D. Hinrichs, and M. Holthaus, Europhys. Lett. \textbf{91}, 10004 (2010).

\bibitem{Iskin}M. Iskin, Phys. Rev. A \textbf{83}, 051606(R) (2011).
\end{thebibliography}
\end{document}